\documentclass[pre,twocolumn,showpacs]{revtex4}
\usepackage{graphicx,amsmath,amssymb}

\def\olcite#1{Ref.~\onlinecite{#1}}
\newcommand{\fig}[1]{Fig.~\ref{#1}}
\newcommand{\avg}[1]{ {\langle #1 \rangle} }

\newcommand{\mypicwidth}{0.8\columnwidth}

\newcommand{\etac}{ \eta_{\rm c} }
\newcommand{\etap}{ \eta_{\rm p} }

\newcommand{\be}{ \beta \epsilon}
\newcommand{\kb}{ k_{\rm B} }
\newcommand{\Pc}{ P(\etac) }

\newcommand{\sigmac}{ \sigma_{\rm c} }
\newcommand{\sigmap}{ \sigma_{\rm p} }

\newcommand{\zc}{ z_{\rm c} }
\newcommand{\zp}{ z_{\rm p} }

\newcommand{\etapr}{ \eta_{\rm p}^{\rm r} }
\newcommand{\rhopr}{ \rho_{\rm p}^{\rm r} }
\newcommand{\etacg}{ \etac^{\rm g} }
\newcommand{\etacl}{ \etac^{\rm l} }

\begin{document}

\title{Simulation and theory of fluid demixing and interfacial tension
       of mixtures of colloids and non-ideal polymers}

\begin{abstract}
 An extension of the Asakura-Oosawa-Vrij model of hard sphere colloids
 and non-adsorbing polymers, that takes polymer non-ideality into
 account through a repulsive stepfunction pair potential between
 polymers, is studied with grand canonical Monte Carlo simulations and
 density functional theory. Simulation results validate previous
 theoretical findings for the shift of the bulk fluid demixing binodal
 upon increasing strength of polymer-polymer repulsion, promoting the
 tendency to mix. For increasing strength of the polymer-polymer
 repulsion, simulation and theory consistently predict the interfacial
 tension of the free colloidal liquid-gas interface to decrease
 significantly for fixed colloid density difference in the coexisting
 phases, and to increase for fixed polymer reservoir packing fraction.
\end{abstract}

\author{R. L. C. Vink}
\affiliation{
 Institut f\"ur Physik, Johannes-Gutenberg-Universit\"at,
 Staudinger Weg 7, D-55099 Mainz, Germany.}

\author{Matthias Schmidt\footnote{Present address:  
    Institut f\"ur Theoretische Physik II,
    Heinrich-Heine-Universit\"at D\"usseldorf, Universit\"atsstra\ss e 1,
    D-40225 D\"usseldorf, Germany.}}
\affiliation{
    Soft Condensed Matter, 
    Debye Institute, Utrecht University, Princetonplein 5,
    3584 CC Utrecht, The Netherlands.}



\date{27 December 2004}

\maketitle

\section{Introduction}

Various different levels of description have been employed in order to
study mixtures of colloidal particles and non-adsorbing globular polymers
\cite{poon02,tuinier03review}. While colloid-colloid interactions are
reasonably well described by those of hard spheres, the effective
interactions between colloids and polymers, as well as that between the
centers of two polymers, are more delicate. The model due to Asakura and
Oosawa \cite{asakura54} and Vrij \cite{vrij76} (AOV), taking the colloids
to be hard spheres and the polymer-polymer interactions to vanish, is
arguably the most simple description of real colloid-polymer mixtures. Its
appeal clearly lies in its simplicity, rather than in very close
resemblance of experimental colloid-polymer systems, enabling detailed
simulation \cite{meijer94, dijkstra99, dijkstra02swet, bolhuis02phasediag,
vink04jcp} and theoretical \cite{gast83, lekkerkerker92, dijkstra99,
schmidt00cip, schmidt02cip} studies of bulk \cite{meijer94, dijkstra99,
bolhuis02phasediag, vink04jcp} and interfacial \cite{brader00,
brader01inhom, dijkstra02swet, brader02swet, brader03swetl,
wessels04codef, wessels04stns} properties. More realistic effective
interactions between the constituent particles can be systematically
obtained by starting at the polymer segment level, and integrating out the
microscopic degrees of freedom of the polymers \cite{jusufi2001,
bolhuis02phasediag}. The resulting colloid-polymer interaction is a
smoothly varying function of distance, and excluded volume between polymer
segments leads to a soft Gaussian-like polymer-polymer repulsion. Such
effective interactions have been used to calculate bulk phase behavior
\cite{dzubiella2002, bolhuis02phasediag}.

In order to retain most of the simplicity of the AOV model, but still to
capture polymer non-ideality, the AOV model was extended using a repulsive
stepfunction pair potential between polymers in \olcite{schmidt03cintp}.
The colloid-polymer interaction was kept as that of hard spheres. This
introduces one additional model parameter, namely the ratio of stepheight
and thermal energy, interpolating between the AOV case when this quantity
vanishes, and the additive binary hard sphere mixture when it becomes infinite.  
Following the hard sphere \cite{rosenfeld89} and AOV cases
\cite{schmidt00cip,schmidt02cip}, the well-defined particle shapes of this
model allowed to obtain a geometry-based density functional theory (DFT)
\cite{evans79,evans92} specifically tailored for this model
\cite{schmidt03cintp}. The theory can, in principle, be applied to
arbitrary inhomogeneous situations, which constitutes an {\em a
posteriori} justification for using these model interactions. The trends
found for the fluid-fluid demixing transition into a colloidal liquid
phase (that is rich in colloids and poor in polymers) and a colloidal gas
phase (that is poor in colloids and rich in polymers), upon increasing the
polymer-polymer interaction strength, demonstrated improved accordance
with the experimental findings of \olcite{ilett95}, as compared to the AOV
case \cite{schmidt03cintp, brader03swetl}, where the DFT predicts the same
phase diagram as the free volume theory \cite{lekkerkerker92}. Furthermore
the extended AOV model was used in the study of the ``floating liquid
phase'' that was found in sedimentation-diffusion equilibrium
\cite{schmidt04aog}. Further discussion of the model and its relation to
the AOV prescription is given in \olcite{schmidt03cintp}.

The aim of the present paper is twofold. First, we want to assess in
detail the accuracy of the DFT of \olcite{schmidt03cintp} by comparing to
results from direct simulations of the extended AOV model. While
phase-coexistence is often studied in the Gibbs ensemble
\cite{panagiotopoulos87bulk}, we instead choose to take advantage of
recently developed methods \cite{virnau2003a, vink04jcp} that rely on the
grand canonical ensemble. Benefits are accurate estimates of the
interfacial tension \cite{binder1982a, vink04jcp} and access to the
critical region \cite{vink04critical}. The binodal we obtain in the
simulations is compared to that from DFT, and good agreement between both
is found. This result strongly supports the original claim
\cite{schmidt03cintp} that a straightforward perturbation theory, taking
the AOV system as the reference state and adding polymer-polymer
interactions in a mean-field like manner, will fail, as the bulk
fluid-fluid binodal from this approach differs markedly from that of the
geometry-based DFT, as discussed in detail in \olcite{schmidt03cintp}.
Instead, even to lowest order in strength of the polymer-polymer
interaction, a non-trivial coupling to the colloid density field needs to
be taken into account; the DFT of \olcite{schmidt03cintp} does this
intrinsically. Our second aim is to study the interfacial tension between
demixed colloidal gas and liquid phases, which is known from experiments
\cite{dehoog99, dehoog01, aarts03swet, aarts04capw}, theory \cite{vrij97,
brader00, brader02swet, brader03swetl} and simulation \cite{vink04jcp,
fortini04tensnew} to be orders of magnitude smaller than that of atomic
substances. While much work has been devoted to the case of
non-interacting polymers \cite{vrij97, brader00, brader02swet,
brader03swetl, wessels04stns, vink04jcp, wessels04codef,
fortini04tensnew}, only quite recent studies addressed the effect of
polymer non-ideality \cite{aarts04cahn, monchojorda2003, monchojorda04}.
Aarts {\it et al.}\ \cite{aarts04cahn} use a square gradient approach
based on the free energy of a free volume theory, to include polymer
interactions \cite{aarts02}, and the mean spherical approximation for the
direct correlation function. They find the gas-liquid interfacial tension
to decrease as compared to the case of non-interacting polymers, when
plotted as a function of the density difference in the coexisting phases.
Similar findings were reported by Moncho-Jord\'a {\it et al.}\
\cite{monchojorda2003}, who also used a square gradient approach, but
based on an effective colloid-colloid depletion potential that reproduces
simulation results accurately \cite{louis02depletion}. Our present study
goes beyond Refs.~\onlinecite{aarts04cahn, monchojorda2003}, and also
beyond the very recent \olcite{monchojorda04}, as we employ a
non-perturbative DFT treating the full two-component mixture of colloids
and interacting polymers. We compare results for the interfacial tension
to our data from direct simulation of the binary mixture. Besides its
intrinsic interest, the interfacial tension is moreover relevant for the
occurrence of capillary condensation in confined systems
\cite{schmidt03capc, schmidt04cape, aarts04codef} as is apparent from a
treatment based on the Kelvin equation \cite{evans87}.

The paper is organized as follows. In Sec.~\ref{SECmodel} we define the
extended AOV model taking into account polymer-polymer repulsion. In
Sec.~\ref{SECmethods} we briefly sketch the theoretical and simulation
techniques. In Sec.~\ref{SECresults} results for fluid-fluid phase
behavior and the interfacial tension are presented, and we conclude in
Sec.~\ref{SECconclusions}.

\section{Model}
\label{SECmodel}

We consider a mixture of hard sphere colloids (species c) of diameter
$\sigmac$, and effective polymer spheres (species p) of diameter
$\sigmap$, that interact via pairwise potentials $V_{ij}(r)$ with $(i,j)
\in ({\rm c,p})$, as function of the center-to-center distance $r$ between
two particles, given as
\begin{eqnarray}
\label{eq:potcc}
  V_{\rm cc}(r) &=&
  \begin{cases}
  \infty & r < \sigmac \\ 
  0 & {\rm otherwise},
  \end{cases}\\
\label{eq:potcp}
  V_{\rm cp}(r) &=&
  \begin{cases}
  \infty & r < (\sigmac + \sigmap)/2 \\
  0 & {\rm otherwise},
  \end{cases}\\
\label{eq:potpp}
  V_{\rm pp}(r) &=&
  \begin{cases}
  \epsilon & r < \sigmap \\
  0 & {\rm otherwise}.
  \end{cases}
\end{eqnarray}
Particle numbers are denoted by $N_i$, and as bulk thermodynamic
parameters we use the packing fractions $\eta_i = \pi \sigma_i^3 N_i /
(6V)$, where $V$ is the system volume. As an alternative to $\etap$, we
use the packing fraction $\etapr = \pi \sigmap^3 \rhopr / 6$ in a
reservoir of pure polymers, interacting via $V_{\rm pp}(r)$ as given
above, that is in chemical equilibrium with the system, with $\rhopr$
being the polymer number density in the reservoir.

The model is characterized by two dimensionless control parameters, namely
the polymer-to-colloid size ratio $q = \sigmap / \sigmac$, and the scaled
strength of polymer-polymer repulsion $\be$, with $\beta=1/(\kb T)$, $\kb$
the Boltzmann constant, and $T$ the absolute temperature. As limiting
cases, the present model possesses the AOV model for $\be=0$, where
polymer-polymer interactions are ideal, and the binary additive hard
sphere mixture in the limit $\be \to \infty$. One can use the parameter
$\be$ to match to a real system at a given thermodynamic statepoint,
polymer type and solvent, by imposing equality of the second
(polymer-polymer) virial coefficients. See \olcite{schmidt03cintp} for an
in-depth discussion of this procedure.

\section{Methods}
\label{SECmethods}

\subsection{Density functional theory}

To investigate bulk and interfacial properties of the present model,
we use the geometrically-based DFT of \olcite{schmidt03cintp} that has
its roots in generalizations of Rosenfeld's fundamental-measure theory
for additive hard sphere mixtures \cite{rosenfeld89}, namely the
treatment of the ``penetrable sphere model'' \cite{schmidt99ps}
(equivalent to the present polymer particles) and the DFT genuinely
developed for the AOV model \cite{schmidt00cip}. The minimization of
the grand potential is carried out with a simple iteration
technique. We refer the reader directly to \olcite{schmidt03cintp} for
all details about the (approximate) Helmholtz free energy functional.

\subsection{Simulation method}

The simulations are carried out in the grand canonical ensemble, where
the fugacities $\zc$ and $\zp$, of colloids and polymers,
respectively, and the total volume $V$ are fixed, while the numbers of
particles, $N_{\rm c}$ and $N_{\rm p}$, are allowed to fluctuate. We
use a rectangular box of dimensions $L \times L \times D$, with
periodic boundary conditions in all three directions, and simulate the
full mixture as defined by the pair potentials given in
Eqs.~(\ref{eq:potcc})-(\ref{eq:potpp}), i.e.~the positional degrees of
freedom of both colloids and polymers are explicitly taken into
account.  Note that {\it asymmetric} binary mixtures are in general
difficult to simulate, and prone to long equilibration times. To
alleviate this problem, we rely on a recently developed cluster
move~\cite{vink04jcp, vink04springer}, that has already been applied
successfully to the standard AOV model \cite{vink04jcp,
vink04springer, vink04codef, vink04critical}. Here we perform the
generalization to $\be>0$ which is straightforward.

During the simulation, we measure the probability $\Pc$ of finding a
certain colloid packing fraction $\etac$. At phase coexistence, the
distribution $\Pc$ becomes bimodal, with two peaks of equal area, one
located at small values of $\etac$ corresponding to the colloidal gas
phase, and one located at high values of $\etac$ corresponding to the
colloidal liquid phase. Typical coexistence distributions for the standard
AOV model can be found in \olcite{vink04jcp}, and our present results
display similar behavior. The equal area rule \cite{muller1995a} implies
that $\int_0^\avg{\etac} \Pc {\rm d}\etac = \int_\avg{\etac}^\infty \Pc
{\rm d}\etac$, with $\avg{\etac}$ the average of the full distribution
$\avg{\etac} = \int_0^\infty \etac \Pc {\rm d}\etac$, where we assume that
$\Pc$ has been normalized to unity $\int_0^\infty \Pc {\rm d}\etac = 1$.  
The packing fraction of the colloidal gas $\etacg$ now follows trivially
from the average of $\Pc$ in first peak $\etacg = 2 \int_0^\avg{\etac}
\etac \Pc {\rm d}\etac$, and similarly for the colloidal liquid $\etacl =
2 \int_\avg{\etac}^\infty \etac \Pc {\rm d}\etac$, where the factors of 2
are a consequence of the normalization of $\Pc$.

The interfacial tension $\gamma$ is extracted from the logarithm of the
probability distribution \mbox{$W \equiv \kb T \ln P(\etac)$.} Note that
$-W$ corresponds to the free energy of the system. Therefore, the height
$F_L$ of the peaks in $W$, measured with respect to the minimum in between
the peaks, equals the free energy barrier separating the colloidal gas
from the colloidal liquid \cite{binder1982a}. $F_L$ may be related to the
interfacial tension $\gamma$ by noting that, at colloid packing fractions
between the peaks $\etacg \ll \etac \ll \etacl$, the system consists of a
colloidal liquid in coexistence with its vapor. A snapshot of the system
in this regime, would reveal a so-called slab geometry, with one region
dense in colloids, and one region poor in colloids, separated by an
interface (because of periodic boundary conditions, two such
interfaces are actually present). If an elongated simulation box with
$D>L$ is used, rather than a cubic box with $D=L$, the interfaces will be
oriented perpendicular to the elongated direction, since this minimizes
the interfacial area, and hence the free energy of the system. The total
interfacial area in the system thus equals $2L^2$ and, following
\olcite{binder1982a}, $\gamma = F_L/(2L^2)$. An additional advantage of
using an elongated simulation box is that interactions between the
interfaces are suppressed. This enhances a flat region in $W$ between the
peaks, which is required for an accurate estimate of the interfacial
tension. In this work, an elongated box of dimensions $D/\sigmac=16.7$ and
$L/\sigmac=8.3$ is used.

Close to the critical point the simulation moves back and forth easily
between the gas and liquid phases, while further away from the critical
point, i.e.~at higher polymer fugacity, the free energy barrier between
the two phases increases. Hence transitions from one to the other phase
become less likely, and the simulation spends most of the time in only one
of the two phases. A crucial ingredient in our simulation is therefore the
use of a biased sampling technique. We employ successive umbrella
sampling, as was recently developed by Virnau and M\"uller
\cite{virnau2003a}, to enable accurate sampling in regions of $\etac$
where $\Pc$, due to the free energy barrier separating the phases, is very
small. In this approach, states (or windows) are sampled successively. In
the first window, the number of colloids is allowed to vary between 0 and
1, in the second window between 1 and 2, and so forth. The number of
polymers is allowed to fluctuate freely in each window. Our sampling
scheme is thus strictly one-dimensional: the bias is put on the colloid
density only. Note that this becomes problematic for very large systems
because of droplet formation. An additional free energy barrier, which
grows with the size of the simulation box, must be crossed before the
transition from the droplet state, to the slab geometry occurs (which is
required if the interfacial tension is to be determined). As pointed out
in Refs.~\onlinecite{macdowell2004, virnau2004}, this additional barrier is
not one in colloid density, but rather in the energy-like parameter, which
for our system would be the polymer density. Hence, for very large
systems, a naive one-dimensional biasing scheme such as described above,
is prone to severe sampling difficulties (more appropriate in this case
would be a two-dimensional sampling scheme in both the colloid {\em and}
the polymer density). For the system size used by us, however, no problems
in obtaining the slab geometry were encountered.

In this work, we simulate up to a colloid packing fraction of $\etac
\approx 0.45$, corresponding to a maximum of around 1000 colloidal
particles. The maximum number of polymers is obtained at low $\etac$.
While the precise number depends on $\etapr$ and $\be$, a value of 3000 is
typical. In each window, ${\cal O}(10^7)$ grand canonical cluster moves
are attempted, of which ${\cal O}(10^5)$ are accepted at low $\etac$, and
${\cal O}(10^3)$ at high $\etac$ (the grand canonical cluster move thus
becomes less efficient with increasing colloid packing fraction). The
typical CPU time investment to obtain a single distribution $\Pc$ is
48~hours on a moderate computer.

\section{Results}
\label{SECresults}

\begin{figure}
\begin{center}

\includegraphics[width=\mypicwidth]{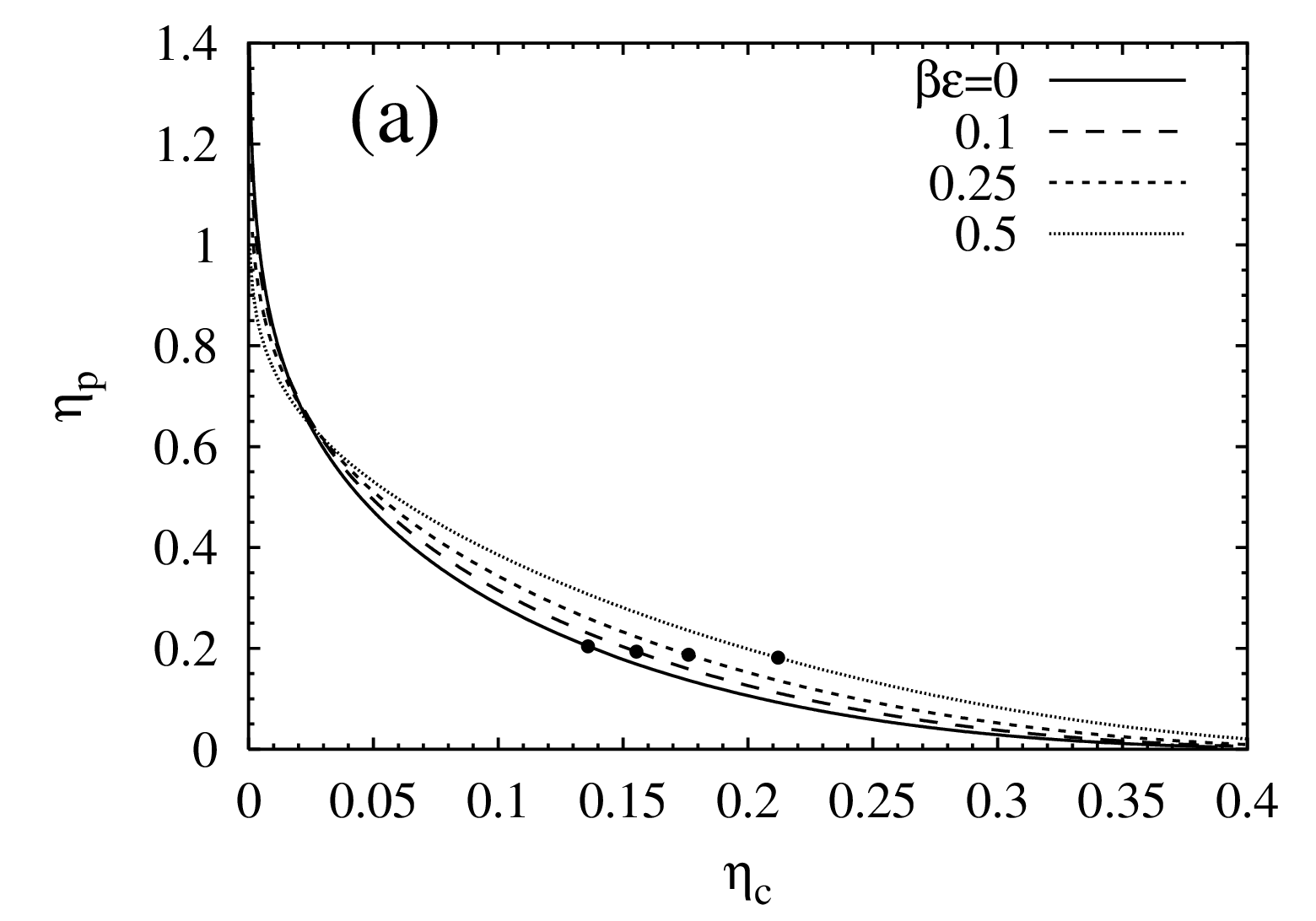}
\includegraphics[width=\mypicwidth]{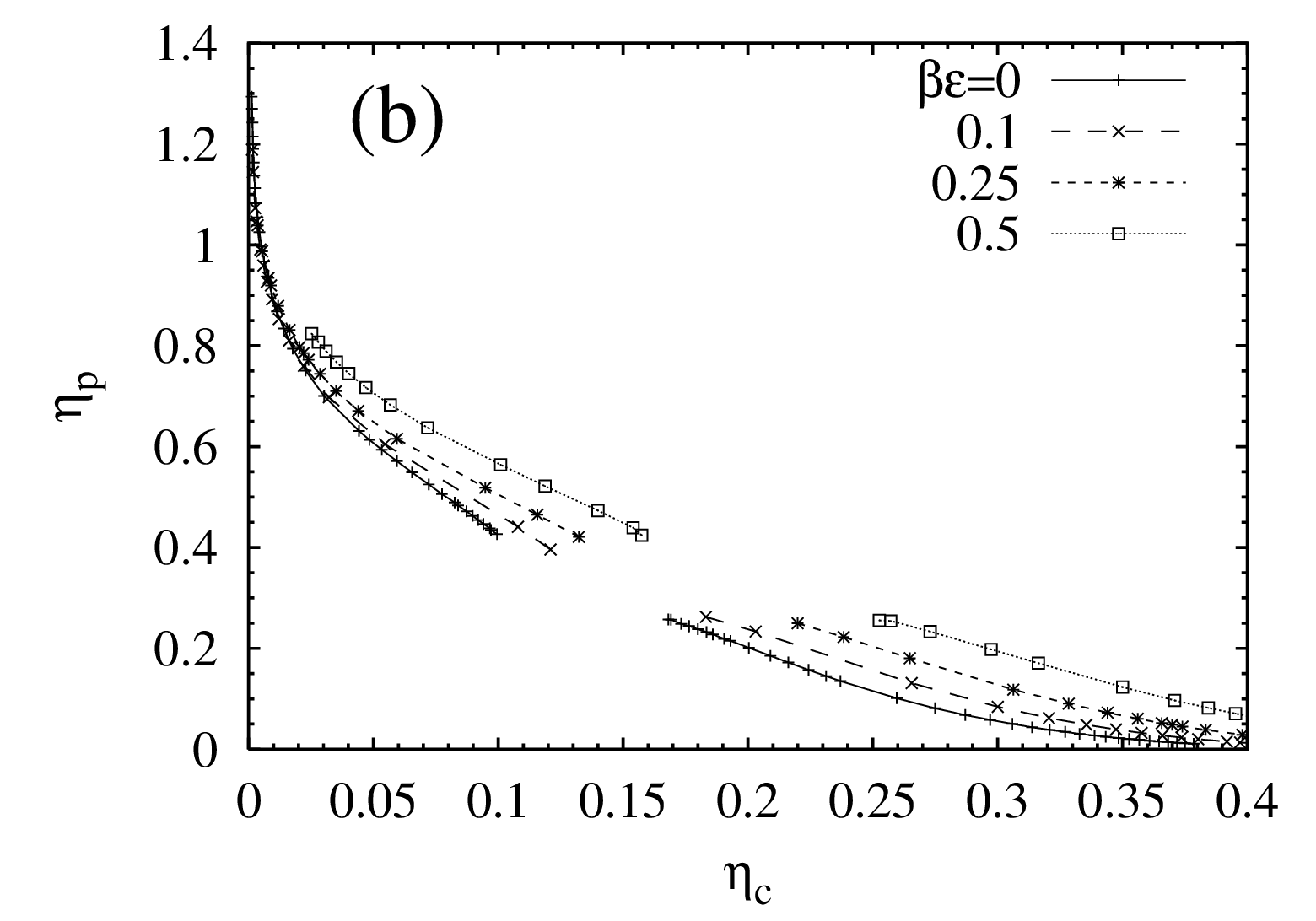}

\caption{Bulk fluid-fluid demixing phase diagram of the extended AOV
model as a function of colloid packing fraction, $\etac$, and polymer
packing fraction, $\etap$, for size ratio $q=0.8$ and increasing strength
of the polymer-polymer repulsion $\be=0,0.1,0.25,0.5$ as indicated. a) The
binodal (lines) and critical point (symbols) as obtained from DFT. The
case $\be=0$ corresponds to the result from free volume theory for the AOV
model. b) The binodal as obtained from simulations (symbols indicate data
points; lines are guides to the eye).}
\label{FIGpdSystem}

\end{center}
\end{figure}

In \fig{FIGpdSystem} we plot results for the bulk fluid-fluid demixing
binodal as obtained from theory and simulation in the
$(\etac,\etap)$-plane, i.e.~the ``system representation''. For the AOV
case (recovered for $\be=0$), the DFT predicts the same bulk free energy
for fluid states, and hence the same demixing binodal, as free volume
theory \cite{lekkerkerker92}. The result is known to compare overall
reasonably well with simulation results for a variety of size ratios $q$,
but to deviate close to the critical point \cite{meijer94,
bolhuis02phasediag, dijkstra02swet, vink04jcp}. For increasing strength of
the polymer-polymer repulsion, the theoretical critical point shifts
toward higher values of $\etac$, and very slightly to lower values of
$\etap$. The accompanying shift of the binodal leads to a growth of the
one-fluid region in the phase diagram, hence polymer-polymer repulsion
tends to promote mixing. The simulation results indicate the same trend,
but show a quantitatively larger shift of the binodal toward higher values
of $\etac$ upon increasing $\be$. Note also the pronounced finite-size
deviations of the simulation data in the vicinity of the critical point.
As a result, the slight decrease in the critical value of $\etap$ with
increasing $\be$, as predicted by DFT, cannot be identified in the
simulation data. To better access the critical region, a finite size
scaling analysis \cite{binder1981a, bruce1992a, kim2003a} would be
required. While such an investigation has been carried out for the
standard AOV model \cite{vink04critical}, it would require extensive
additional simulations for the extended model, which is beyond the scope
of the present work.

\begin{figure}
\begin{center}

\includegraphics[width=\mypicwidth]{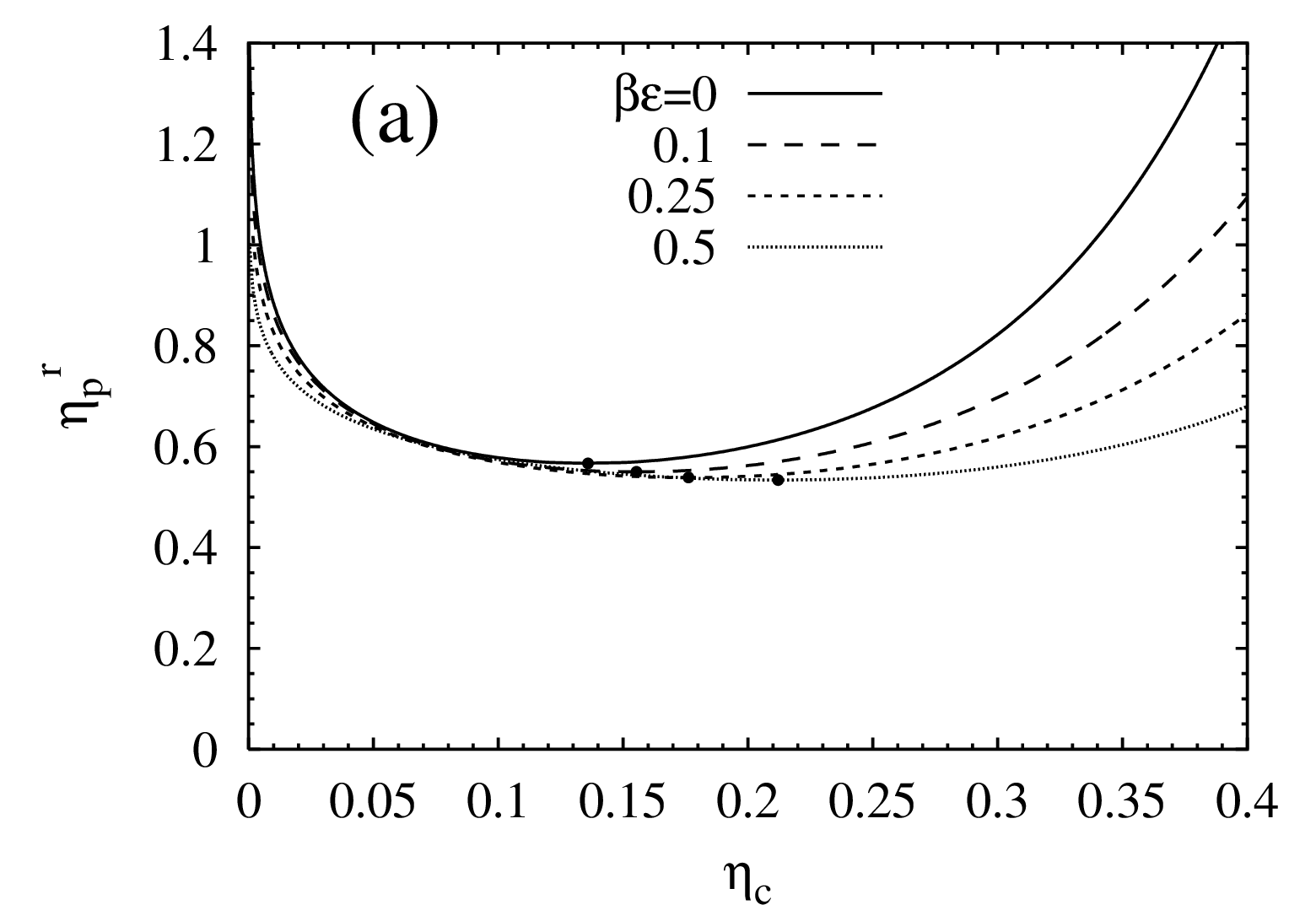}
\includegraphics[width=\mypicwidth]{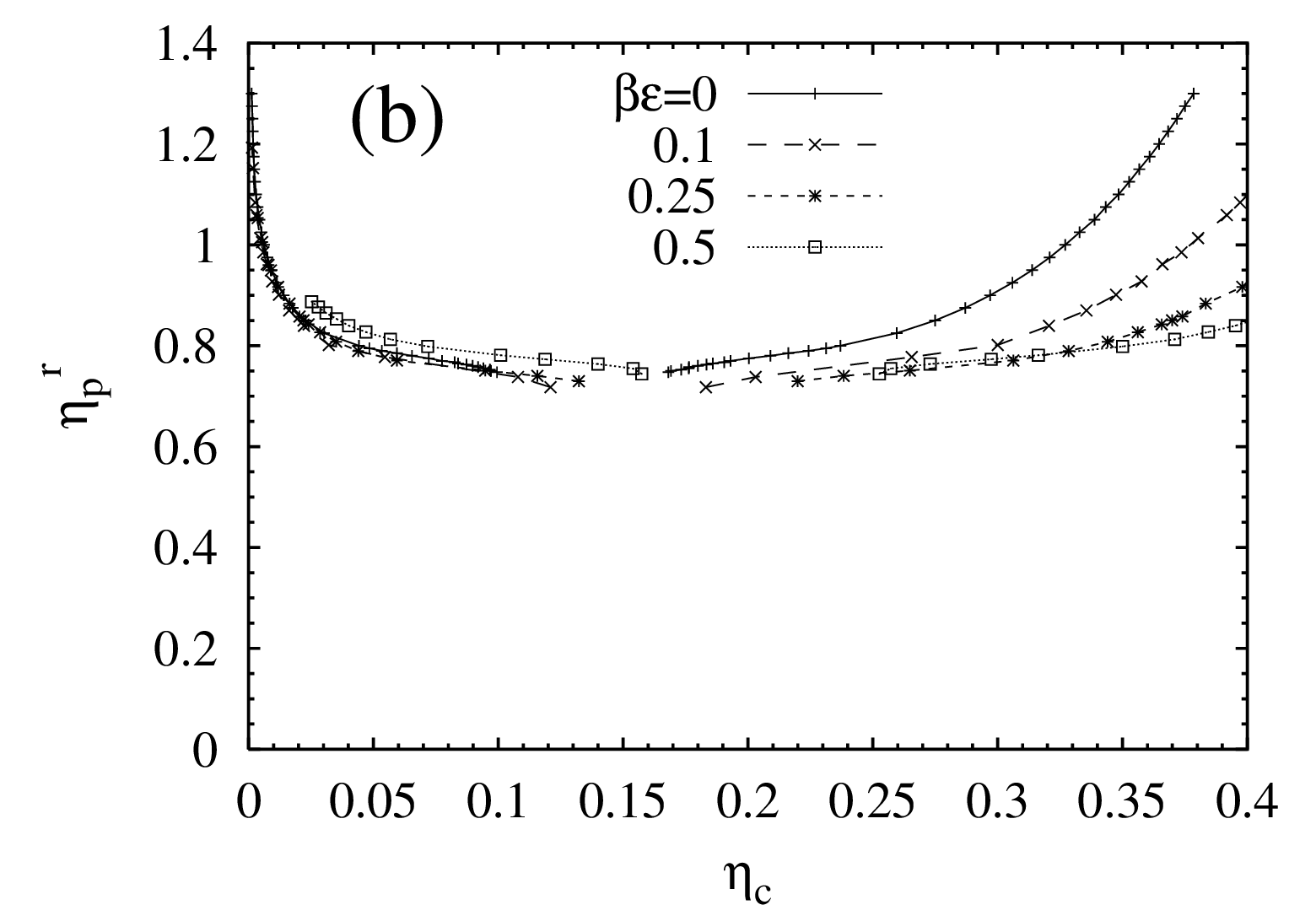}
    
\caption{The analogue of \fig{FIGpdSystem}, but as a function of colloid
   packing fraction, $\etac$, and polymer reservoir packing fraction
   $\etapr$ in a reservoir of interacting polymers.}
\label{FIGpdReservoir}

\end{center}
\end{figure}

When plotted in the $(\etac,\etapr)$-plane or ``reservoir
representation'', the theoretical results display a similar shift of the
binodal toward larger values of $\etac$, and considerable broadening of
the coexistence region, see \fig{FIGpdReservoir}. This implies that at a
given value of $\etapr$ above the critical point, increasing the
polymer-polymer repulsion leads to a larger density difference between the
coexisting phases. The broadening of the coexistence region is strikingly
confirmed by the simulation data. We again observe an overall shift of the
binodals toward higher values of $\etac$. In accordance with findings for
the standard AOV model \cite{meijer94, bolhuis02phasediag, dijkstra02swet,
vink04jcp}, the critical value of $\etapr$ obtained by DFT underestimates
the simulation value. This is related to the universality class of the AOV
model, which is that of the 3D Ising model \cite{vink04critical}, and
hence, close to criticality, deviations from a mean-field treatment like
the present DFT are to be expected.

\begin{figure}
\begin{center}

\includegraphics[width=\mypicwidth]{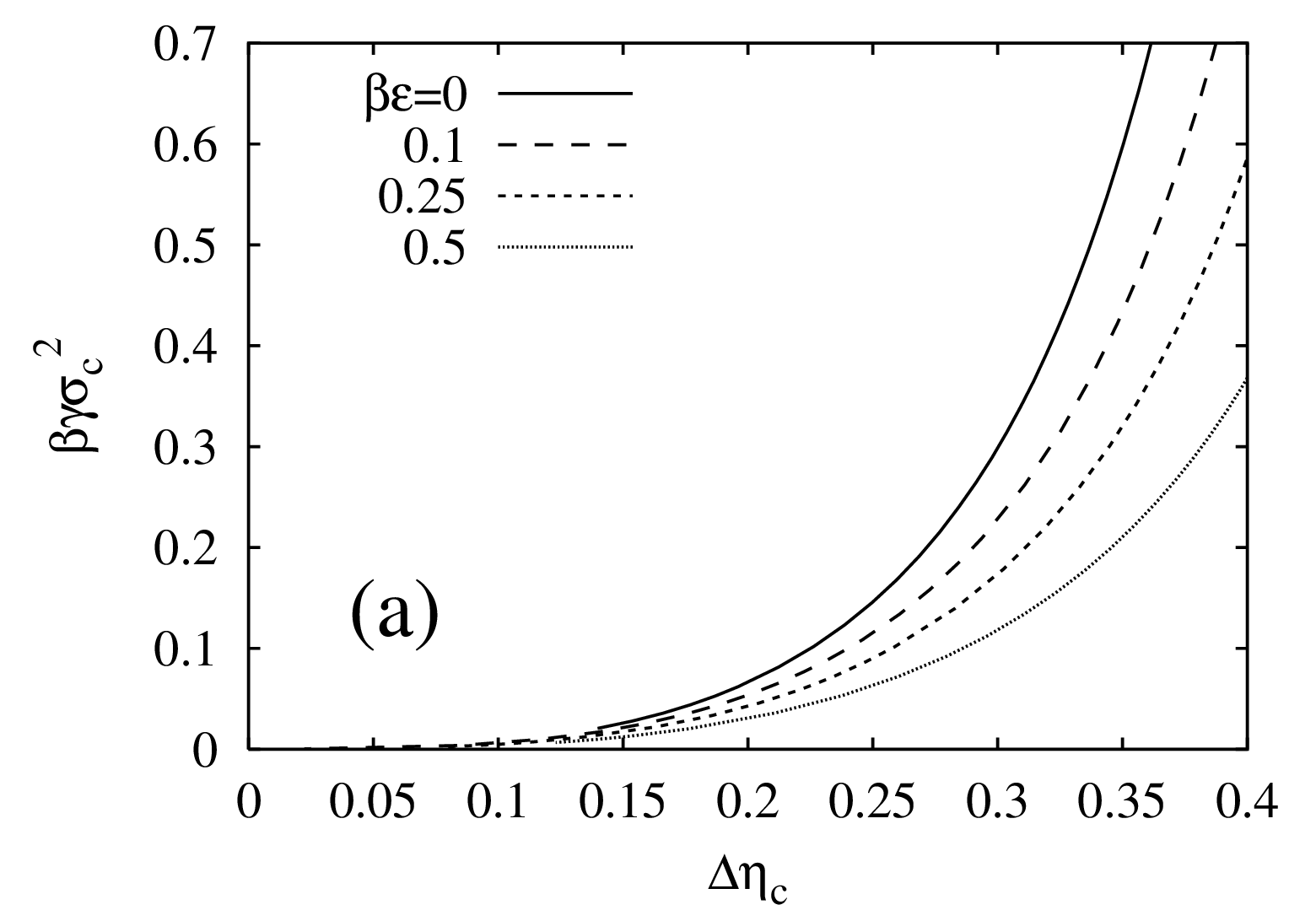}
\includegraphics[width=\mypicwidth]{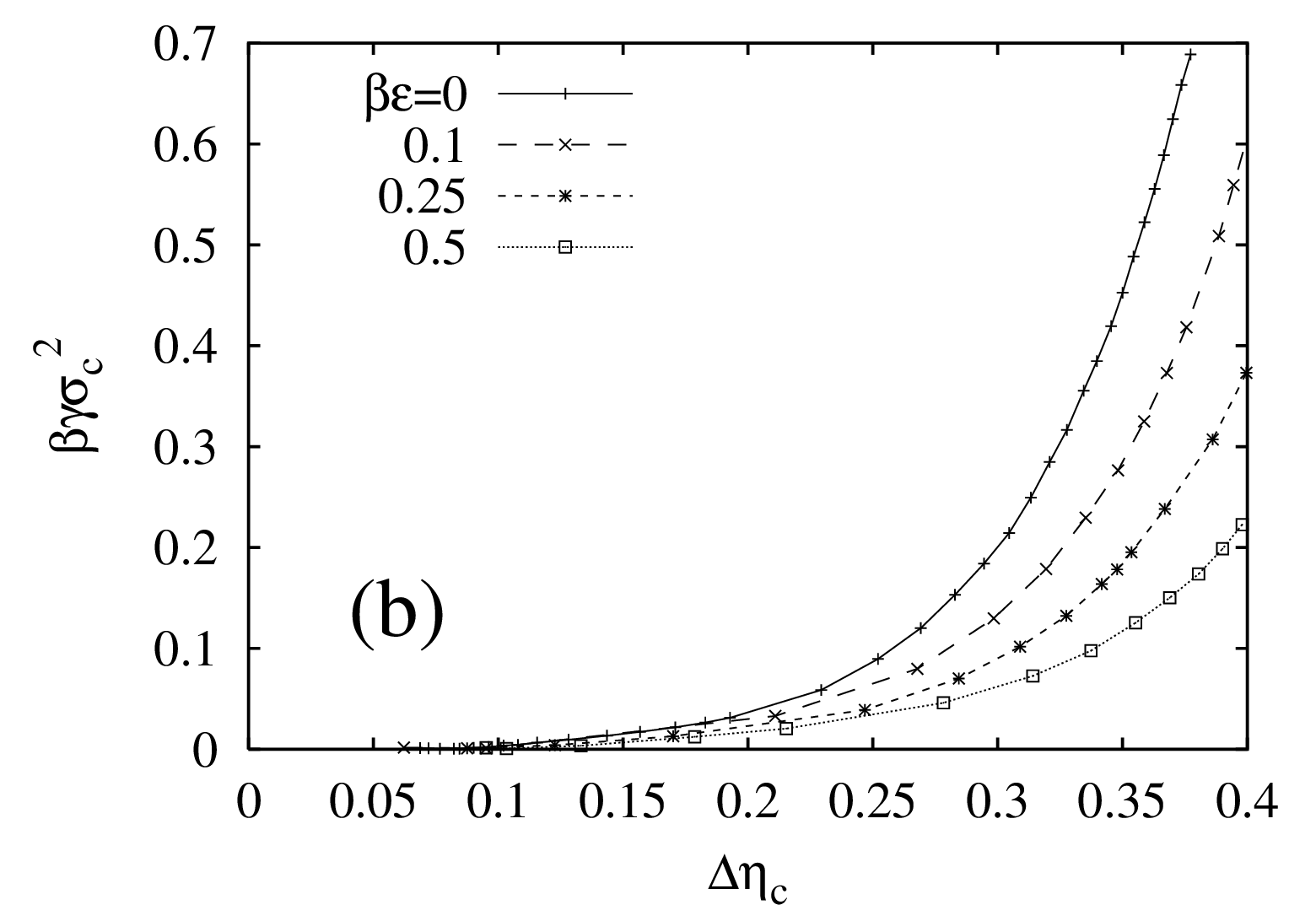}
    
\caption{Dimensionless interfacial tension $\beta \gamma \sigmac^2$,
   of the free colloidal gas-liquid interface, as a function of the
   difference in colloid packing fractions of the coexisting states,
   $\Delta \etac = \etacl - \etacg$. The size ratio is $q=0.8$, and
   the strength of the polymer-polymer repulsion equals
   $\be=0,0.1,0.25,0.5$ as indicated. a) Results from DFT; b) results
   from simulation.}
\label{FIGgammaDeltaEta}

\end{center}
\end{figure}

In \fig{FIGgammaDeltaEta}, we show results for the gas-liquid interfacial
tension, $\gamma$, plotted as a function of the difference between the
colloid packing fraction in the coexisting phases, $\Delta \etac = \etacl
- \etacg$. Both DFT and simulation indicate that the interfacial tension
decreases markedly upon increasing $\be$, in accordance with
square-gradient treatments \cite{aarts04cahn, monchojorda2003}. Note that
this change is not due to interfacial contributions alone, but also to the
pronounced changes in the location of the bulk demixing binodal discussed
above. The decrease might seem reasonable based on the profound increase
in the density of the coexisting liquid upon increasing $\be$, see again
the phase diagram in reservoir representation (\fig{FIGpdReservoir}). As
is apparent from \fig{FIGpdReservoir}, a given value of $\Delta \etac$
corresponds to a reduction of $\etapr$ at coexistence upon increasing
$\be$. We believe that this decrease induces the observed decrease in
$\gamma$.

\begin{figure}
\begin{center}

\includegraphics[width=\mypicwidth]{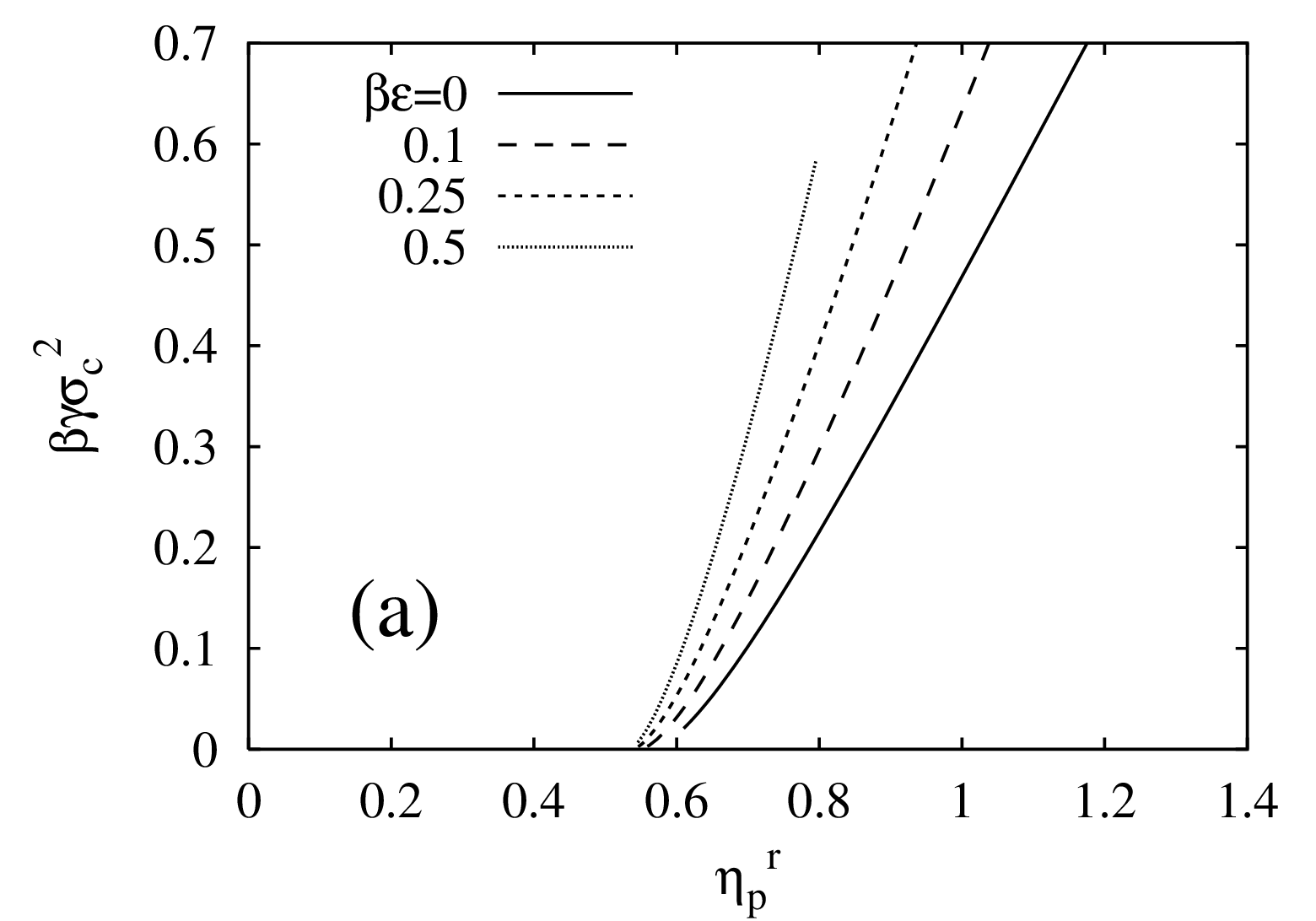}
\includegraphics[width=\mypicwidth]{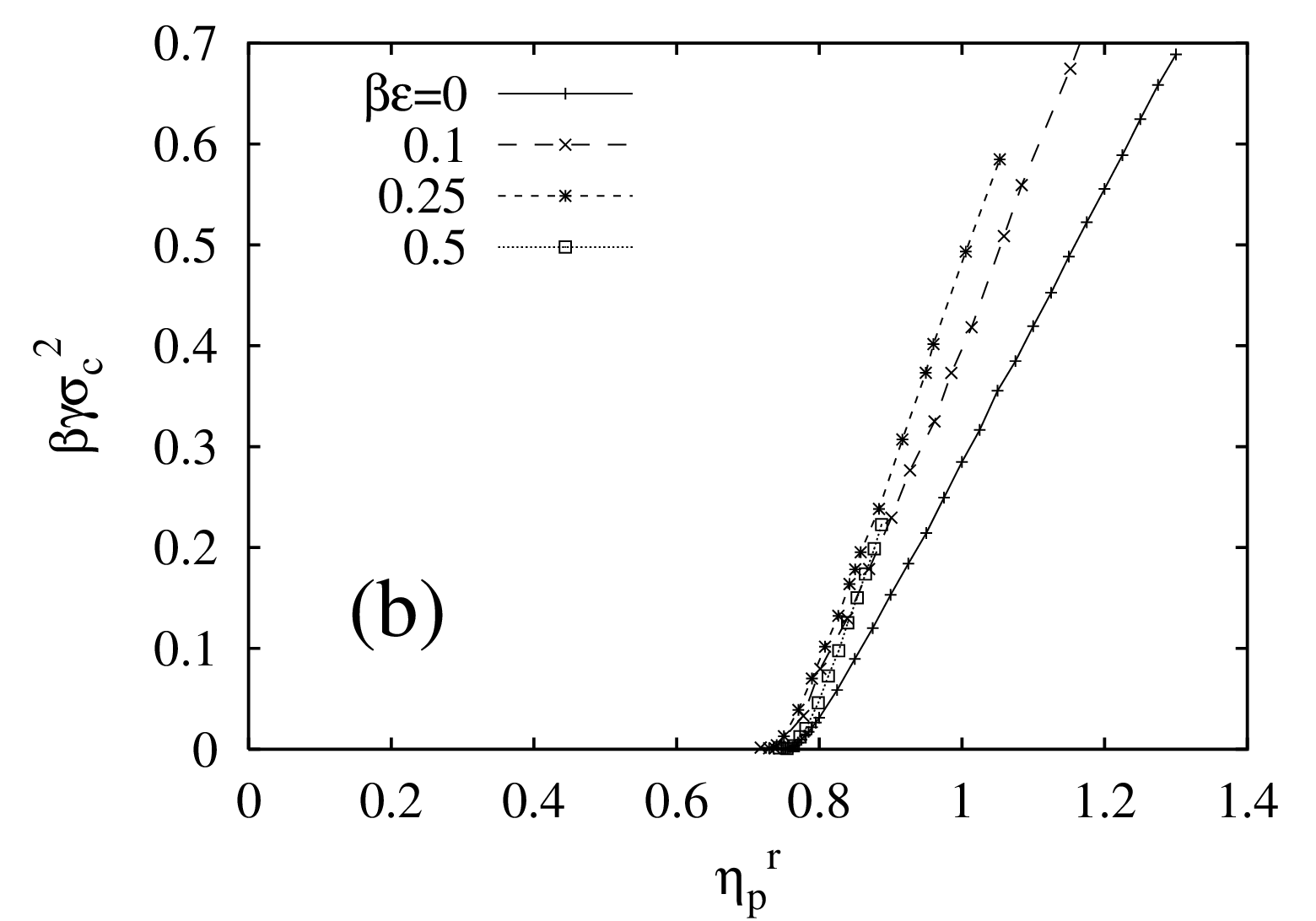}

\caption{The analogue of \fig{FIGgammaDeltaEta}, but as a function of the
   polymer reservoir packing fraction $\etapr$. a) Results from DFT; b) 
   results from simulation.}
\label{FIGgammaReservoir}

\end{center}
\end{figure}

Alternatively, comparing the interfacial tension as a function of
$\etapr$ thus reveals an {\em increase} with increasing $\be$, see the
upper panel of \fig{FIGgammaReservoir} which shows the DFT result. The
lower panel of \fig{FIGgammaReservoir} shows the corresponding
simulation data. The increase in the interfacial tension with
increasing $\be$ is confirmed up to $\be=0.25$, but not for
$\be=0.5$. The simulation data for the latter case, however, should be
treated with some care, as in grand canonical simulations, the
quantity $\etapr$ does not play the role of a direct control
parameter, which rather the polymer fugacity, $\zp$,
does. Consequently, conversion to the reservoir representation
requires an additional step, namely the determination of $\etapr$ as a
function of $\zp$. With the exception of $\be=0$, in which case
$\etapr = \pi \sigmap^3 \zp / 6$ holds trivially, this conversion
introduces an additional statistical uncertainty. \fig{FIGpdReservoir}
shows that, for $\be=0.5$, the binodal has become very flat, so even
small uncertainties in $\etapr$ imply rather large uncertainties in
$\Delta \etac$. Hence, for very flat binodals, the reservoir
representation is not convenient in simulations (more reliable in this
case is the system representation, see \fig{FIGgammaDeltaEta}, which,
as a benefit, is experimentally more relevant). A second point is that
the Monte Carlo cluster move becomes less efficient with increasing
$\etac$ and $\be$ (see Ref.~\onlinecite{vink04jcp}) and this will also
adversely affect the data (especially for $\be=0.5$, since then both
$\be$ and $\etacl$ are substantial). Therefore, we conclude that the
shift of the simulation data for $\be=0.5$ in \fig{FIGgammaReservoir}
most likely reflects a simulation artifact.

\section{Conclusions}
\label{SECconclusions}

In conclusion, we have investigated the effect of polymer-polymer
repulsion on the fluid-fluid demixing phase behavior and on the
(colloidal) liquid-gas interface of a model colloid-polymer mixture. We
have used a simplistic pair potential between polymers, given by a
repulsive stepfunction, to extend the standard AOV model to cases of
interacting polymers. Grand canonical Monte Carlo simulations of the full
mixture demonstrate the reasonable accuracy of the theory, with a tendency
to quantitatively underestimate the shifts in the bulk fluid-fluid
demixing binodal, and the gas-liquid interfacial tension, upon increasing
strength of the polymer-polymer repulsion. The present study demonstrates
the usefulness of the model as such, as it clearly displays previously
found features due to polymer non-ideality. This offers ways to study
further interesting inhomogeneous situations, like the wetting properties
at substrates. Such investigations could test the robustness of the
results obtained for the surface phase behavior at a hard wall using ideal
polymers \cite{brader02swet, brader03swetl, dijkstra02swet, wessels04stns,
wessels04codef}.

\acknowledgments

We thank J\"urgen Horbach, Peter Virnau, Marcus M\"{u}ller, Kurt Binder,
Marjolein Dijkstra and Andrea Fortini for many useful and inspiring
discussions. The work of MS is part of the research program of the {\em
Stichting voor Fundamenteel Onderzoek der Materie} (FOM), which is
financially supported by the {\em Nederlandse Organisatie voor
Wetenschappelijk Onderzoek} (NWO). Support is acknowledged by the SFB-TR6
``Physics of colloidal dispersions in external fields'' of the
\emph{Deutsche Forschungsgemeinschaft} (DFG). RLC also acknowledges
generous allocation of computer time on the JUMP at the Forschungszentrum
J\"{u}lich GmbH.

\clearpage


\begin{thebibliography}{10}

\bibitem{poon02}
W.~C.~K. Poon, J. Phys.:\ Condensed Matter {\bf 14},  R859  (2002).

\bibitem{tuinier03review}
R. Tuinier, J. Rieger, and C.~G. {de Kruif}, Adv. Colloid Interface Sci. {\bf
  103},  1  (2003).

\bibitem{asakura54}
S. Asakura and F. Oosawa, J. Chem. Phys. {\bf 22},  1255  (1954).

\bibitem{vrij76}
A. Vrij, Pure and Appl. Chem. {\bf 48},  471  (1976).

\bibitem{meijer94}
E.~J. Meijer and D. Frenkel, J. Chem. Phys. {\bf 100},  6873  (1994).

\bibitem{dijkstra99}
M. Dijkstra, J.~M. Brader, and R. Evans, J. Phys.:\ Condensed Matter {\bf 11},
  10079  (1999).

\bibitem{vink04jcp}
R.~L.~C. Vink and J. Horbach, J. Chem. Phys. {\bf 121},  3253  (2004).

\bibitem{bolhuis02phasediag}
{P. G. Bolhuis, A. A. Louis, J. P. Hansen}, Phys. Rev. Lett. {\bf 89},  128302
  (2002).

\bibitem{dijkstra02swet}
M. Dijkstra and R. {van Roij}, Phys. Rev. Lett. {\bf 89},  208303  (2002).

\bibitem{gast83}
A.~P. Gast, C.~K. Hall, and W.~B. Russell, J. Coll. Int. Sci. {\bf 96},  251
  (1983).

\bibitem{lekkerkerker92}
H.~N.~W. Lekkerkerker, W.~C.~K. Poon, P.~N. Pusey, A. Stroobants, and P.~B.
  Warren, Europhys. Lett. {\bf 20},  559  (1992).

\bibitem{schmidt00cip}
M. Schmidt, H. L{\"o}wen, J.~M. Brader, and R. Evans, Phys. Rev. Lett. {\bf
  85},  1934  (2000).

\bibitem{schmidt02cip}
M. Schmidt, H. L{\"o}wen, J.~M. Brader, and R. Evans, J. Phys.:\ Condensed
  Matter {\bf 14},  9353  (2002).

\bibitem{brader00}
J.~M. Brader and R. Evans, Europhys. Lett. {\bf 49},  678  (2000).

\bibitem{brader01inhom}
J.~M. Brader, M. Dijkstra, and R. Evans, Phys. Rev. E {\bf 63},  041405
  (2001).

\bibitem{brader02swet}
J.~M. Brader, R. Evans, M. Schmidt, and H. L{\"o}wen, J. Phys.:\ Condensed
  Matter {\bf 14},  L1  (2002).

\bibitem{brader03swetl}
J.~M. Brader, R. Evans, and M. Schmidt, Mol. Phys. {\bf 101},  3349  (2003).

\bibitem{wessels04codef}
P.~P.~F. Wessels, M. Schmidt, and H. L{\"o}wen, J. Phys.:\ Condensed Matter
  {\bf 16},  S4169  (2004).

\bibitem{wessels04stns}
P.~P.~F. Wessels, M. Schmidt, and H. L{\"o}wen, J. Phys.:\ Condensed Matter
  {\bf 16},  L1  (2004).

\bibitem{jusufi2001}
A. Jusufi, J. Dzubiella, C.~N. Likos, C. {von Ferber}, and H. L{\"o}wen, J.
  Chem. Phys. {\bf 116},  9518  (2002).

\bibitem{dzubiella2002}
J. Dzubiella, C.~N. Likos, and H. L{\"o}wen, J. Chem. Phys. {\bf 116},  9518
  (2002).

\bibitem{schmidt03cintp}
M. Schmidt, A.~R. Denton, and J.~M. Brader, J. Chem. Phys. {\bf 118},  1541
  (2003).

\bibitem{rosenfeld89}
Y. Rosenfeld, Phys. Rev. Lett. {\bf 63},  980  (1989).

\bibitem{evans79}
R. Evans, Adv. Phys. {\bf 28},  143  (1979).

\bibitem{evans92}
R. Evans,  in {\em Fundamentals of Inhomogeneous Fluids}, edited by D.
  Henderson (Dekker, New York, 1992), Chap.~3, p.\ 85.

\bibitem{ilett95}
S.~M. Ilett, A. Orrock, W.~C.~K. Poon, and P.~N. Pusey, Phys. Rev. E {\bf 51},
  1344  (1995).

\bibitem{schmidt04aog}
M. Schmidt, M. Dijkstra, and J.~P. Hansen, Phys. Rev. Lett. {\bf 93},  088303
  (2004).

\bibitem{panagiotopoulos87bulk}
A.~Z. Panagiotopoulos, Mol. Phys. {\bf 61},  813  (1987).

\bibitem{virnau2003a}
P. Virnau and M. M{\"u}ller, J. Chem. Phys. {\bf 120},  10925  (2003).

\bibitem{binder1982a}
K. Binder, Phys. Rev. A {\bf 25},  1699  (1982).

\bibitem{vink04critical}
R.~L.~C. Vink, J. Horbach, and K. Binder, accepted for publication in Phys.
  Rev. E {\bf 70} (2004).

\bibitem{dehoog99}
E.~H.~A. {de Hoog} and H.~N.~W. Lekkerkerker, J. Phys. Chem. B {\bf 103},  5274
   (1999).

\bibitem{dehoog01}
E.~H.~A. {de Hoog} and H.~N.~W. Lekkerkerker, J. Phys. Chem. B {\bf 105},
  11636  (2001).

\bibitem{aarts03swet}
D.~G. A.~L. Aarts, J.~H. {van der Wiel}, and H.~N.~W. Lekkerkerker, J. Phys.:\
  Condensed Matter {\bf 15},  S245  (2003).

\bibitem{aarts04capw}
D.~G. A.~L. Aarts, M. Schmidt, and H.~N.~W. Lekkerkerker, Science {\bf 304},
  847  (2004).

\bibitem{vrij97}
A. Vrij, Physica A {\bf 235},  120  (1997).

\bibitem{fortini04tensnew}
A. Fortini, M. Dijkstra, M. Schmidt, and P.~P.~F. Wessels, submitted
 to Phys. Rev. E.

\bibitem{aarts04cahn}
D.~G. A.~L. Aarts, R.~P.~A. Dullens, H.~N.~W. Lekkerkerker, D. Bonn, and R.
  {van Roij}, J. Chem. Phys. {\bf 120},  1973  (2004).

\bibitem{monchojorda2003}
A. Moncho-Jord\'a, B. Rotenberg, and A.~A. Louis, J. Chem. Phys. {\bf 119},
  12667  (2003).

\bibitem{monchojorda04}
A. Moncho-Jord\'a, J. Dzubiella, J.~P. Hansen, and A.~A. Louis,
  cond-mat/0411282.

\bibitem{aarts02}
D.~G. A.~L. Aarts, R. Tuinier, and H.~N.~W. Lekkerkerker, J. Phys.:\ Condensed
  Matter {\bf 14},  7551  (2002).

\bibitem{louis02depletion}
A.~A. Louis, P.~G. Bolhuis, E.~J. Meijer, and J.~P. Hansen, J. Chem. Phys. {\bf
  117},  1893  (2002).

\bibitem{schmidt03capc}
M. Schmidt, A. Fortini, and M. Dijkstra, J. Phys.:\ Condensed Matter {\bf 48},
  S3411  (2003).

\bibitem{schmidt04cape}
M. Schmidt, A. Fortini, and M. Dijkstra, J. Phys.:\ Condensed Matter {\bf 16},
  S4159  (2004).

\bibitem{aarts04codef}
D.~G. A.~L. Aarts and H.~N.~W. Lekkerkerker, J. Phys.:\ Condensed Matter {\bf
  16},  S4231  (2004).

\bibitem{evans87}
R. Evans and U. {Marini Bettolo Marconi}, J. Chem. Phys. {\bf 86},  7138
  (1987).

\bibitem{schmidt99ps}
M. Schmidt, J. Phys.:\ Condensed Matter {\bf 11},  10163  (1999).

\bibitem{vink04springer}
R.~L.~C. Vink,  in {\em Computer Simulation Studies in Condensed Matter Physics
  XVIII}, edited by D.~P. Landau, S.~P. Lewis, and H.~B. Schuettler (Springer,
  Berlin, 2004).

\bibitem{vink04codef}
R.~L.~C. Vink and J. Horbach, J. Phys.:\ Condensed Matter {\bf 16},  S3807
  (2004).

\bibitem{muller1995a}
M. M{\"u}ller and N.~B. Wilding, Phys. Rev. E {\bf 51},  2079  (1995).

\bibitem{macdowell2004}
L.~G. MacDowell, P. Virnau, M. M{\"u}ller, and K. Binder, J. Chem. Phys. {\bf
  120},  5293  (2004).

\bibitem{virnau2004}
P. Virnau, L.~G. MacDowell, M. M{\"u}ller, and K. Binder,  in {\em High
  Performance Computing in Science and Engineering 2004}, edited by S. Wagner,
  W. Hanke, A. Bode, and F. Durst (Springer, Berlin, 2004), p.\ 125.

\bibitem{binder1981a}
K. Binder, Z. Phys. B {\bf 34},  119  (1981).

\bibitem{bruce1992a}
A.~D. Bruce and N.~B. Wilding, Phys. Rev. Lett. {\bf 68},  193  (1992).

\bibitem{kim2003a}
Y.~C. Kim, M.~E. Fisher, and E. Luijten, Phys. Rev. Lett. {\bf 91},  65701
  (2003).

\end{thebibliography}

\clearpage


\end{document}